\documentclass[11pt,paper]{JHEP}

\usepackage{amssymb}
\usepackage{amsmath}
\usepackage{amsfonts}
\usepackage{cite}

\allowdisplaybreaks[1]

\def\makeatletter{\catcode`\@=11}
\makeatletter
\def\mathbox#1{\hbox{$\m@th#1$}}%
\def\math@ccstyles#1#2#3#4#5#6#7{{\leavevmode
      \setbox0\mathbox{#6#7}%
      \setbox2\mathbox{#4#5}%
      \dimen@ #3%
      \baselineskip\z@\lineskiplimit#1\lineskip\z@
      \vbox{\ialign{##\crcr
             \hfil \kern #2\box2 \hfil\crcr
             \noalign{\kern\dimen@}%
             \hfil\box0\hfil\crcr}}}}
\def\mathaccstyles{\math@ccstyles\maxdimen}
\def\maththroughstyles{\math@ccstyles{-\maxdimen}}
\def\unitmatrixDT%
 {\maththroughstyles{.45\ht0}\z@\displaystyle {\mathchar"006C}\displaystyle 1}
\def\leftrightarrowfill{$\mathsurround0pt \mathord\leftarrow
       \mkern-6mu\cleaders\hbox{$\mkern-2mu \mathord- \mkern-2mu$}\hfill
       \mkern-6mu \mathord\rightarrow$}

\def\overleftrightarrow#1{\vbox{\ialign{##\crcr
        \leftrightarrowfill\crcr\noalign{\kern-1pt\nointerlineskip}%
        $\hfil\displaystyle{#1}\hfil$\crcr}}}

\numberwithin{table}{section}

\title{About the $S^3$ Group-manifold Reduction of Einstein
Gravity}
\author{Rom\'an Linares-Romero\\
Departamento de F\'{\i}sica, Universidad Aut\'onoma Metropolitana Iztapalapa, \\
San Rafael Atlixco 186, c.p. 09340, M\'exico D.F., M\'exico.\\
     E-mail: \email{lirr@xanum.uam.mx}}

\abstract{We exhibit a new consistent group-manifold reduction of
pure Einstein gravity in the vielbein formulation when the
compactification group manifold is $S^3$. The novel feature in the
reduction is to exploit the two 3-dimensional Lie algebras that
$S^3$ admits. The first algebra is introduced into the
group-manifold reduction in the standard way through the
Maurer-Cartan 1-forms associated to the symmetry of the general
coordinate transformations. The second algebra is associated to the
linear adjoint group and it is introduced into the group-manifold
reduction through a local transformation in the internal tangent
space. We discuss the characteristics of the resulting
lower-dimensional theory and we emphasize the novel results
generated by the new group-manifold reduction. As an application of
the reduction we show that the lower-dimensional theory admits a
domain wall solution which upon uplifting to the higher-dimension
results to be the self-dual (in the non-vanishing components of both
curvature and spin connection) Kaluza-Klein monopole. This
discussion may be relevant in the dimensional reductions of
$M$-theory, string theory and also in the Bianchi cosmologies in
four dimensions.}


\begin{document}

\section{Introduction\label{Intro}}

Unification of gravitation and gauge theories in a
higher-dimensional theory is one of the most interesting and
attractive ideas in physics. If the truly fundamental theory is
higher-dimensional, then one should be interested in all its
predictions and consequences, including solutions. In particular, if
one hopes that the 4-dimensional world may be described by the
fundamental theory, it is necessary to fully understand the
mechanisms (dimensional reductions) to extract lower-dimensional
physics from the higher-dimensional theory.

In the dimensional reduction one starts with the curvature scalar
(and other possible fields) as the Lagrangian in $D+d$ dimensions.
Next one assumes general covariance and supposes that because of
some dynamical mechanism, the background manifold $M_{D+d}$ is the
direct product of two manifolds: the $d$-dimensional internal
compactification manifold $M_d$ and the background manifold $M_D$ of
the resulting $D$-dimensional theory. It turns out that dimensional
reductions can be divided into two types
\cite{Cvetic:2003jy,Gibbons:2003gp}. The first type of reductions
consider a quotient space $G/H$ as the internal manifold and are
called {\it coset reductions} or Pauli reductions
\cite{Straumann:2000zc}. In this paper we shall not discuss this
kind of reductions any further.

The second type of reductions are based on the assumption that the
parameterizations for the metric and other higher-dimensional fields
are invariant under a $d$-dimensional simply transitive acting group
of isometries in the internal space. These reductions include both
the original Kaluza-Klein reduction on $S^1$ in which the group of
isometries is $U(1)$ \cite{Kaluza:1921tu,Klein:1926tv}, and the {\it
group-manifold reductions} where the group of isometries is the left
action $(G_d)_L$ of the group manifold $G_d$. Actually, the metric
on the orbit space of the group $G_d$ is bi-invariant, i.e. it has
$(G_d)_L \times (G_d)_R$ as its isometry group, but as DeWitt
indicated \cite{DeWitt}, a fully consistent reduction involves a
metric which is merely left-invariant. In the literature, the
group-manifold reductions are sometimes called DeWitt reductions
\cite{DeWitt} and sometimes Scherk-Schwarz reductions
\cite{Scherk:1979zr}. All these reductions are {\it consistent}
because the group invariance of the parametrization ensures that
every solution of the lower-dimensional equations of motion
corresponds to a solution of the higher-dimensional equations of
motion.

Well studied examples of group-manifold reductions are the ones that
consider a $(D+3)$-dimensional Hilbert-Einstein action as the
starting theory and a 3-dimensional internal space invariant under a
3-dimensional group of isometries. In three dimensions the isometry
groups are characterized locally by the eleven inequivalent
3-dimensional Lie algebras \cite{Bianchi,Estabrook}. In this paper
we are interested in dimensional reductions on Bianchi type IX group
manifolds, which are defined to be manifolds with an $SO(3)$ or
$SU(2)$ isometry group acting transitively on 3-surfaces. Locally
both isometry groups are characterized by the same Lie algebra,
although topologically they are different, $SU(2)$ is the double
covering of $SO(3)$ and they correspond to $S^3$ and $\mathbb{R}P^3$
respectively ($\mathbb{R}P^3$ is $S^3$ with antipodal points
identified).

If one performs the group-manifold reduction of pure Einstein
gravity considering to the metric as the basic field, all the
geometrical information needed in the reduction is contained in the
space-time symmetry. The lower-dimensional theory obtained in this
way is an Einstein-Maxwell-scalars gauged theory where the isometry
group of the internal space becomes the gauge group of both the
Maxwell fields and the scalars of the internal coset space. In the
case of Bianchi type IX group manifolds the gauge group of the
lower-dimensional theory is either $SO(3)$ or $SU(2)$. For this
reason in the literature this dimensional reduction is called
$S^3=SU(2)$ group-manifold reduction.

Although the metric formulation is appropriate for pure gravity, the
presence of spinors requires the introduction of a longer set of
variables. These are the vielbein fields which describe local
orthonormal Lorentz frames at each space-time point and with respect
to which the spinors are defined. In order to treat the
group-manifold reduction in the general case is therefore important
to perform the reduction of the gravitational sector using the
vielbein fields as basic variables. In this formulation gravity has
two different local symmetries, the space-time symmetry and the
tangent Lorentz symmetry. The standard group-manifold reduction of
gravity in the vielbein formulation only exploits the space-time
symmetry \cite{Scherk:1979zr}.

The standard Bianchi type IX group-manifold reduction in the
vielbein formulation has been applied to many theories, among them,
to the 4-dimensional Einstein gravity \cite{Ellis:1969vb}, the
($D+3$)-dimensional Einstein gravity, the bosonic string
\cite{Cvetic:2003jy}, the 11-dimensional supergravity
\cite{Salam:1985ft} and to the 10-dimensional supergravity
\cite{Chamseddine:1997nm,Chamseddine:1999uy}.

The purpose of this paper is to exhibit a {\it new consistent} way
to perform the $S^3$ group-manifold reduction of pure Einstein
gravity in the vielbein formulation. The novel feature in the
reduction is to exploit the two 3-dimensional Lie algebras that the
group manifold $S^3$ admits. One of the groups is introduced into
the group-manifold reduction in the standard way through the
Maurer-Cartan 1-forms associated to the symmetry of the general
coordinate transformations. The another group is dictated by the
symmetry of the internal tangent space and it is introduced into the
group-manifold reduction through the linear adjoint group
\cite{Ellis:1969vb,Jantzen}. The role this latter group plays in the
spatial topology of the internal manifold has been discussed in
\cite{Luciani:1978zv,Ashtekar:wa,Graham:1994qs,Koike:rb}. We shall
show that the introduction of the adjoint matrix $\Lambda$ in the
parametrization of the vielbein leads to non trivial differences in
the group-manifold reduction. These differences are: a) A new term
in the components of the spin connection with two internal indices
and b) an additional term in the covariant derivative of the
internal ``triad".

We have two main motivations to study this new group-manifold
reduction. The first one is related to some results in the context
of the 4-dimensional $N=1$ supergravity. Starting with the
supersymmetry constraints for full supergravity and then only
considering Bianchi type IX homogeneous configurations of the metric
an fields, the authors of \cite{Graham:1994qs} find: a) There are
two distinct definitions of homogeneity and therefore two different
Ans\"atze for the fermion fields and dreibein. One possibility
corresponds to the isometry group $SO(3)$ and the another one to
$SU(2)$. b) The different Ans\"atze for the fermion fields depend on
whether the spinor components may have the same or opposite sign at
antipodal points of the spatial 3-manifold. c) The different
Ans\"atze for the dreibein differ by a orthogonal matrix $\Lambda$
that depends of the internal coordinates. d) The expressions and
solutions of the supersymmetry constraints depend on the Ans\"atze
used for the dreibein and the Rarita-Schwinger field. For the zero
fermion state, the standard definition of homogeneity (without
considering $\Lambda$) gives rise to a wormhole state
\cite{D'Eath:1993up}, whereas the definition of homogeneity
involving the matrix $\Lambda$ leads to a Hartle-Hawking state. It
is important to stress that although $\Lambda$ is introduced in the
Ans\"atze for the fields in \cite{Graham:1994qs}, the group-manifold
reduction is not performed. In this paper we wish to close this gap.

The second motivation is to get a better understanding of recent
results concerning domain wall solutions of 8-dimensional gauged
supergravities
\cite{Hernandez:2002fb,AlonsoAlberca:2003jq,Bergshoeff:2003ri} and
the relation of these solutions to the classification of
3-dimensional compactification manifolds, both, locally (Bianchi
classification \cite{Bianchi}) and globally (Thurston classification
\cite{Thurston}). The different 8-dimensional gauged supergravities
\cite{Salam:1985ft,Bergshoeff:2003ri} arise from group-manifold
reductions of the 11-dimensional supergravity \cite{Cremmer:1978km}
over the different 3-dimensional compactification manifolds. In
particular the 8-dimensional gauged supergravity of Salam and Sezgin
\cite{Salam:1985ft} admits 1/2 BPS domain wall solutions which upon
uplifting to eleven dimensions become purely gravitational solutions
with metrics of the form $\mathbb{R}^{6,1} \times M_4$. In this case
$M_4$ are the self-dual metrics (in both curvature and spin
connection) of Belinsky-Gibbons-Page-Pope. It happens that the three
equations obtained by require a self-dual spin connection for $M_4$
are exactly the same that the ones obtained by require a 1/2 BPS
domain wall solution to the 8-dimensional transformation rules for
the dilatinos. From the 11-dimensional point of view the uplifted
solutions are 1/2 BPS except for an special case which uplift to
11-dimensional flat space and hence becomes fully supersymmetric. A
disturbing fact is that the Kaluza-Klein monopole
\cite{Sorkin:1983ns,Gross:1983hb} is also a purely gravitational 1/2
BPS solution of the 11-dimensional supergravity, however by reducing
it applying the standard group-manifold reduction, the supersymmetry
in eight dimensions becomes fully broken. This happens because in
the frame of the standard group-manifold reduction the Kaluza-Klein
monopole which has the form $\mathbb{R}^{6,1} \times M_4$ does not
have self-dual spin connection for $M_4$.

In this paper we shall exhibit the similarities and differences
obtained from the two consistent group-manifold reductions of the
$(D+3)$-dimensional Hilbert-Einstein action in the vielbein
formulation on Bianchi type IX group manifolds. As an application of
the reductions, we study the domain wall solutions of the resulting
$D$-dimensional theory and its uplifting to $(D+3)$-dimensions. We
shall get the solutions at the level of the first order differential
equations that emerge from the self-duality condition of the
non-vanishing components of the higher-dimensional spin connection.
From the $(D+3)$-dimensional point of view these solutions are of
the form $\mathbb{R}^{D-2,1} \times M_4$. It is a well known fact
that by performing the standard group-manifold reduction, the system
of equations obtained by require self-dual spin connection in $M_4$
results to be the ``Belinsky-Gibbons-Page-Pope" first order system
\cite{Belinskii:1978}. By performing the new group-manifold
reduction we show that the self-duality condition of the spin
connection leads to the ``Atiyah-Hitchin" first order system
\cite{Atiyah:1985dv}. As a consequence the $D$-dimensional theory
admits a domain wall solution which upon uplifting to $D+3$
dimensions leads to the self-dual (in both the curvature and the
spin connection of $M_4$) Kaluza-Klein monopole. A preliminary
presentation of some of our results can be found in \cite{Linares}.
Here we extend the discussion of the new group-manifold reduction
and we show its relation with other well known results in the
literature.

The outline of the paper is as follows. In section \ref{E-H} we
perform the $S^3$ group-manifold reduction of the
$(D+3)$-dimensional Hilbert-Einstein action. We start in \ref{sgct}
summarizing the discussion about the group-manifold reduction of the
general coordinate transformations given in \cite{Scherk:1979zr}. In
\ref{tangent} we introduce the adjoint matrix $\Lambda$ associated
to the group manifold $S^3$ whereas in \ref{anzatse} we introduce
the new parametrization of the vielbein and we compare it with the
parametrization of the standard group-manifold reduction. We perform
the group-manifold reduction of the spin connection and the action
in \ref{action}. In section \ref{domainw} we obtain the domain wall
solutions of the reduced action. We start analyzing the solutions to
the second order differential equations of motion in \ref{secondor}
and in \ref{firstor} we discuss the domain wall solutions from the
point of view of the self-duality condition of the spin connection.
We conclude the section in \ref{supot} writing down the first-order
Bogomol'nyi equations associated to the lower-dimensional action.
Our conclusions are given in section \ref{conclusions}. In appendix
\ref{Lg3d} we explicitly construct the different quantities involved
in the reduction.

\section{$S^3$ group-manifold reduction \label{E-H}}

This section will focus on computing the new $S^3$ group-manifold
reduction of the $(D+3)$-dimensional Hilbert-Einstein action. In the
vielbein formulation, Einstein gravity has two local symmetries, the
general coordinate symmetry and the tangent Lorentz symmetry. We
shall consider the vielbein parametrization in terms of
lower-dimensional fields that involves besides the usual
3-dimensional Lie algebra associated to the general coordinate
transformations of the internal space another 3-dimensional Lie
algebra associated to the local tangent symmetry
\cite{Graham:1994qs}.

In the following discussion we assume a ($D+3$) split of the ($D+3$)
space-time coordinates $x^{\hat{\mu}} = (x^\mu, z^\alpha)$ where
$\mu=\{ 0,1,\ldots ,D-1 \}$ are the indices of the $D$-dimensional
space-time and $\alpha=\{ 1,2,3 \}$ are the indices of the internal
coordinates. The corresponding flat indices of the tangent space are
denoted by $\hat{a} = (a, m)$. The group indices are also denoted
with the letters $m, n,\dots $,. We work in the conventions of
\cite{Bergshoeff:2003ri}.

\subsection{General coordinate transformations\label{sgct}}

In the vielbein formalism, the $(D+3)$-dimensional Hilbert-Einstein
action
\begin{equation}
 S=\int d^{D+3} \hat{x} \ \hat{e} \  \hat{{\cal R}} (\hat {\omega}) \, ,
\end{equation}
is invariant under the general coordinate transformations
\begin{equation}
 \delta_{\hat{\bf K}} \hat {e}_{\hat \mu}{}^{\hat a}
 = {\cal L}_{\hat{\bf K}}\hat {e}_{\hat \mu}{}^{\hat a}
 = \hat{K}^{\hat \nu}\, \partial_{\hat \nu}
 \hat {e}_{\hat \mu}{}^{\hat a}+\partial_{\hat \mu} \hat{K}^{\hat \nu} \,
 \hat {e}_{\hat \nu}{}^{\hat a} \, .
\label{gct}
\end{equation}
As usual, $\hat {e}$ is the determinant of the vielbein, $\hat{{\cal
R}}$ the Ricci scalar, $\hat {\omega}$ the spin connection and
${\cal L}_{\hat{\bf K}}$ denotes the Lie derivative along the
infinitesimal vector field parameters $\hat{\bf K}$.

As it has been pointed out in \cite{Scherk:1979zr}, the
group-manifold reduction is specified by choosing the internal
coordinate dependence of the parameters $\hat{K}^{\hat{\mu}}(x,z)$.
If they are taken as
\begin{equation}
\hat{K}^{\mu}(x,z)=K^\mu(x),\hspace{1cm} \hat{K}^\alpha(x,z)
 = K^m (x) (U^{-1}(z))_m{}^{\alpha} \, ,
\label{kc}
\end{equation}
where $U_\alpha{}^m(z)$ are $GL(3,\mathbb{R})$ matrices which can be
interpreted as the components of the left invariant Maurer-Cartan
1-forms $\sigma^m \equiv dz^\alpha U_\alpha{}^m(z)$, an arbitrary
3-dimensional Lie algebra can be extracted out of the group of
general coordinate transformations in $(D+3)$-dimensions. The
algebra of general coordinate transformations
\begin{equation}
 [\delta_{\hat {K}_1}, \delta_{\hat {K}_2} ]=\delta_{\hat {K}_3} \hspace{1cm}
 \hbox{where} \hspace{1cm}
\hat {K}^{\hat{\mu}}_3(x,z) = 2 \hat {K}^{\hat{\nu}}_{[2}(x,z)\,
\partial_{\hat{\nu}}\hat {K}^{\hat{\mu}}_{1]}(x,z) \, ,
\end{equation}
gives origin to three different possibilities in $D$-dimensions.
First, the algebra of two space-time transformations with parameters
$K^{\mu}_{1}(x)$ and $K^{\mu}_{2}(x)$ gives a new space-time
transformation with parameter $K^{\mu}_3(x) = 2 K^{\nu}_{[2}(x)\,
\partial_{\nu} K^{\mu}_{1]}(x)$ indicating that the theory has
general coordinate transformations in the $D$-dimensional
space-time. Second, the commutator of a space-time transformation
with parameter $K^{\mu}_{1}(x)$ and an internal transformation with
parameter $K^m_{2}(x)$ gives a new internal transformation with
parameter $K^m_3(x) = K^{\mu}_{1}(x)
\partial_{\mu}K^m_{2}(x)$ which means that the parameters of an
internal transformation are space-time scalars. Finally, the
commutator of two internal transformations with parameters
$K^m_{1}(x)$ and $K^m_{2}(x)$ produces a new internal transformation
with parameter $K^p_{3}(x)=f_{mn}{}^p K^m_{1}(x) K^n_{2}(x)$ where
\begin{equation}
 f_{m n}{}^{p} = -2(U^{-1}(z))_{m}{}^{\alpha} (U^{-1}(z))_{n}{}^{\beta}\,
 \partial_{[\alpha} U_{\beta]}{}^p(z)\, ,
\label{MC}
\end{equation}
are the structure constants of the 3-dimensional Lie group $G_3$,
whose Lie algebra $\mathfrak{g}_{\,_{3}}$ is given by
\begin{equation}
[{\bf K}_m,{\bf K}_n]=f_{mn}{}^{p}{\bf K}_p , \label{algk}
\end{equation}
and the $f_{mn}{}^p$'s satisfy the Jacobi identity
$f_{[mn}{}^qf_{p]q}{}^r=0$.

After apply the group-manifold reduction \cite{DeWitt,Scherk:1979zr}
the simply transitive 3-dimensional Lie algebra (\ref{algk}) becomes
the algebra of the gauged group in the lower-dimensional theory. It
turns out that in three dimensions there exist eleven different ways
to choose the structure constants \cite{Bianchi,Estabrook} and
therefore eleven group-manifold reductions \cite{Ellis:1969vb}.
Among them we are interested in the Bianchi type IX group-manifold
reductions for which the structure constants are $f_{mn}{}^p =
\varepsilon_{mnq}\delta^{pq}$. Bianchi type IX metrics are defined
to be manifolds with either an $SO(3)$ or $SU(2)$ isometry group
acting on 3-surfaces. Topologically the group $SO(3)$ is the
projective space $\mathbb{R}P^3$ whilst the group $SU(2)$ is $S^3$.
The projective space $\mathbb{R}P^3$ results from $S^3$ by
identifying pairs of antipodal points. Explicitly the vectors ${\bf
K}_m$ are given by
\begin{eqnarray}
{\bf K}_1 & = & \frac{\cos z^3}{\cos z^2} \partial_1 + \sin z^3
\partial_2 -
\frac{\cos z^3 \sin z^2}{\cos z^2} \partial_3 \, ,\nonumber \\
{\bf K}_2 & = & -\frac{\sin z^3}{\cos z^2} \partial_1 + \cos z^3
\partial_2 +
\frac{\sin z^3 \sin z^2}{\cos z^2} \partial_3 \, , \\
{\bf K}_3& = &  \partial_3 \, \nonumber .
\end{eqnarray}
with
\begin{equation*}
0 \leq z^1 \leq  2\pi, \hspace{1cm} -\frac{\pi}{2}  \leq  z^2 \leq
\frac{\pi}{2},
\end{equation*}
and
\begin{eqnarray}
0 \leq & z^3 & \leq  2\pi, \hspace{1cm} \mbox{if $G_3$ is $SO(3)$} , \label{paramso3}\\
0 \leq & z^3 & \leq  4\pi, \hspace{1cm} \mbox{if $G_3$ is $SU(2)$}
\label{intsu2}.
\end{eqnarray}

\subsection{The adjoint matrix\label{tangent}}

If we perform the group-manifold reduction of pure gravity using the
metric as the basic field, all the geometrical information is
codified in the general coordinate transformations. However, if we
perform the group-manifold reduction using the vielbein as the basic
field, we have two different possibilities for the properties of the
fermion components (even if we do not introduce fermions
explicitly). They can have either the same or opposite sign at
antipodal points of the spatial 3-manifold \cite{Graham:1994qs}.
These two different possibilities must be codified in the symmetry
of the internal tangent space and reflected in geometrical
quantities such as the spin connection.

The novel ingredient of the new group-manifold reduction in the
vielbein formulation is the introduction of the matrix $\Lambda(z)$,
which is taken in the {\it adjoint representation} of the
3-dimensional Lie algebra (\ref{algk}) of the previous section
\cite{Ellis:1969vb,Jantzen}. The mathematical properties of the
linear adjoint group are discussed in detail in \cite{Jantzen}.
Here, we only point out some properties of the adjoint matrix and in
the appendix \ref{Lg3d} we give explicit representations of them.

The adjoint matrix $\Lambda(z)$ is determined by the equation
\begin{equation}
 \Lambda(z) = e^{z^1 R_1} e^{z^2 R_2} e^{z^3 R_3},
\end{equation}
where the constant matrices $R_m$ are the generators of the Lie
algebra ${gl}(3,\mathbb{R})$ in its natural basis $\{ {\bf e}_p{}^n
\}$, and are given by the {\it adjoint representation} of the
parameters of the internal general coordinate transformations,
$R_m=f_{mn}{}^p{\bf e}_p{}^n=ad_{{\bf K}}({\bf K}_m)$. They satisfy
the $AD(SU(2))=SO(3)$ Lie algebra
\begin{equation}
[R_m,R_n]= f_{mn}{}^p R_p. \label{algr}
\end{equation}
The relation between the adjoint matrix and the left invariant
Maurer-Cartan 1-forms is given by ${\bf \Lambda}^{-1}d{\bf \Lambda}
= \sigma^m R_m$. This expression is very important in the
group-manifold reduction because it relates the two different
matrices that contain the geometrical information of the internal
manifold. In components the relation reads
\begin{equation}
 (R_{m})_n{}^{p} = (U^{-1}(z))_{m}{}^{\alpha }
 (\Lambda^{-1}(z))_n{}^q \partial_{\alpha}\Lambda_{q}{}^{p}(z) \,
 .
 \label{LM}
\end{equation}
It can be shown that the adjoint matrix is orthogonal and also that
$\det \Lambda = 1$ . These properties indicate that $\Lambda(z)$ can
be considered as a rotation matrix in the internal tangent space.

\subsection{Parametrization of the vielbein\label{anzatse}}

The next step in the group-manifold reduction is to make a suitable
parametrization of the group-invariant vielbein in terms of
lower-dimensional fields. The parametrization includes internal
coordinates dependence dictated by the symmetries of the theory. It
turns out that for Bianchi type IX group-manifold reductions there
are two possibilities to do it. The first possibility has been
extensively discussed and only considers internal coordinates
dependence through the components $U(z)$ of the Maurer-Cartan
1-forms \cite{Scherk:1979zr}, which are related to the symmetry of
the general coordinates transformations. Throughout this paper we
refer to this reduction as the standard group-manifold reduction.
The second possibility considers besides $U(z)$ a new dependence on
the internal coordinates through the adjoint matrix $\Lambda(z)$,
which is related to the symmetry of the internal local tangent
space. Throughout the paper we refer to this reduction as the ``new"
group-manifold reduction. The parametrization of the vielbein for
the latter possibility is
\begin{align}
  {\hat e}_{\hat \mu}{}^{\hat a}(x,z) = \left(
\begin{array}{cc}
 e^{c_1 \varphi (x)} e_{\mu}{}^{a}(x)  &
 e^{c_2 \varphi (x)} A_{\mu}{}^{\alpha}(x,z) \, L_{\alpha}{}^{p}(x,z) \\
         &                                          \\
      0  & e^{c_2 \varphi (x)} L_{\alpha}{}^{p}(x,z) \\
\end{array}
\right) \, , \label{ansatzmat}
\end{align}
where $c_1$ and $c_2$ are constants whose values are
$c_1=-\sqrt{3}/\sqrt{2(D+1)(D-2)}$ and $c_2 =-{c_1 (D-2)}/{3}$
\footnote{The values of $c_1$ and $c_2$ ensure that the reduction of
the Hilbert-Einstein action yields a pure Hilbert-Einstein term in
$D$-dimensions, with no pre-factor involving the scalar $\varphi$,
and that $\varphi$ has a canonically normalized kinetic term in
$D$-dimensions.}. The $A_{\mu}$'s are gauge fields and
$L_\alpha{}^p(x,z)$ is a $3 \times 3$ matrix whose internal
coordinates dependence are given by
\begin{eqnarray}\label{Apar}
A_{\mu}{}^{\alpha}(x,z) & = &
A_{\mu}{}^m(x)\,(U^{-1}(z))_m{}^{\alpha}
\, , \label{Apar} \\
 L_{\alpha}{}^{p}(x,z) & = & U_{\alpha}{}^m (z)\, L_m{}^n(x) \,
\Lambda_n{}^p(z) \, . \label{nanzatse}
\end{eqnarray}
The vielbein parametrization of the standard group-manifold
reduction can be obtained from (\ref{nanzatse}) replacing the
adjoint matrix $\Lambda$ by the identity matrix. The parametrization
of the vielbein can be rewritten in the shorter form
\begin{eqnarray}\label{eampar}
{\hat e}^a(x,z) & = &  e^{c_1 \varphi(x)} e^a(x), \\
{\hat e}^m(x,z) & = & e^{c_2 \varphi(x)}(A^n(x)+\sigma^n(z))
L_n{}^p(x)\Lambda_p{}^m(z) \equiv e^p(x,z)\Lambda_p{}^m(z).
\label{eint}
\end{eqnarray}
The group-manifold reduction works out because the internal
coordinate dependence can be factored out in any geometrical
quantity due to the fact that it always appears in one of the two
possible combinations (\ref{MC}) or (\ref{LM}). This means for
instance that upon reduction $\delta
L_\alpha{}^p(x,z)=U_\alpha{}^m(z) \, \delta
L_m{}^n(x)\Lambda_n{}^p(z)$.

From the $D$-dimensional point of view under a space-time
transformation $K^{\mu}(x)$, the fields $\varphi(x)$ and
$L_m{}^n(x)$ transform as scalars whilst $e_{\mu}{}^a(x)$ and
$A_\mu{}^m(x)$ transform as vectors, and under an internal
transformation $K^m(x)$, the fields $e_{\mu}{}^a(x)$ and
$\varphi(x)$ do not transform whilst the fields $A_\mu{}^m(x)$ and
$L_m{}^n(x)$ transform in the following way
\begin{eqnarray}
 \delta A_{\mu}{}^m (x)& = & (\partial_\mu K^m-A_\mu{}^n f_{np}{}^m
 K^p) , \\
 \delta L_{m}{}^{n}(x) & = & (f_{mp}{}^qL_q{}^n + L_m{}^q(R_p)_q{}^n) K^p .  \label{dtrans}
\end{eqnarray}
The conclusion from the first equation is that the $A_\mu{}^m$'s are
gauge potentials for the corresponding gauge group $G_3$ whose Lie
algebra is (\ref{algk}). In (\ref{dtrans}) we have the first
consequence due to the introduction of the adjoint matrix $\Lambda$
in the parametrization of the vielbein. Additional to the standard
term $f_{mp}{}^qL_q{}^n$ originated by the equation (\ref{MC}) and
related to the gauging of the $SU(2)$ Lie algebra (\ref{algk}), we
have the new term $L_m{}^q(R_p)_q{}^n$ originated by the equation
(\ref{LM}) and related to the $SO(3)$ Lie algebra (\ref{algr}).
These two terms shall be part of the covariant derivative of the
scalar fields $L_{m}{}^{n}$.

Using the vielbein parametrization
(\ref{ansatzmat})-(\ref{nanzatse}) we can rewrite the 11-dimensional
interval in the way
\begin{equation}
ds^2=e^{2c_1 \varphi}g_{\mu \nu} dx^{\mu} dx^{\nu} - e^{2c_2
\varphi} {\cal M}_{mn}(dx^\mu A_\mu{}^m +\sigma^m)(dx^\nu A_\nu{}^n
+\sigma^n),
\end{equation}
where
\begin{equation}\label{intmetric}
{\cal M}_{mn}(x)\equiv -L_m{}^p(x)L_n{}^q(x)\, \eta_{pq}\, .
\end{equation}
In general $L_m{}^n(x,z)$ describes the 6-dimensional
$GL(3,\mathbb{R})/SO(3)$ scalar coset of the internal space and can
be interpreted as the internal ``triad". It transforms under a
global $GL(3,\mathbb{R})$ acting from the left and a local $SO(3)$
symmetry acting from the right. By a gauge fixing of the SO(3)
symmetry, is possible to find an explicit representative of it
\cite{Bergshoeff:2003ri,AlonsoAlberca:2003jq}. The matrix ${\cal
M}_{mn}(x)$ is the $SO(3)$ invariant metric of the internal manifold
and it is parameterized by the same scalars. In particular, for the
case of Bianchi type IX group manifolds the Lie algebra (\ref{algk})
corresponds to the algebra of the maximal compact subgroup of
${GL(3,\mathbb{R})}$ i.e. $SO(3)$ (it is also the Lie algebra of
$SU(2)$) and the scalar coset $L_m{}^n$ is 5-dimensional.

At this point is clear that if we consider quantities that only
depend of the internal metric ${\cal M}_{mn}$, the local tangent
symmetry is irrelevant. This is not the case if we consider
geometrical quantities whose definition is given in terms of the
triad $L_m{}^n$ such as the spin connection. The main result of this
paper is to realize that is possible to consider the adjoint matrix
$\Lambda(z)$ in a {\it consistent} group-manifold dimensional
reduction scheme.

Upon reduction the independence of the internal coordinates
$z^\alpha$ is guaranteed because it is factored out in any quantity.
Explicitly, if $\hat T(x,z)$ is a $(D+3)$-dimensional field, upon
reduction for each index $\alpha$ or $m$ that it contains, the
internal dependence appears in one of the following ways
\begin{eqnarray}
\hat T^\alpha(x,z) & = & t^m(x)(U^{-1}(z))_m{}^\alpha \ ,
\hspace{0.5cm}
\hat T_\alpha(x,z) = U_\alpha{}^m(z) t_m(x) \, ,\\
\hat T^m(x,z)& = & t^n(x)\Lambda_n{}^m(z) \, , \hspace{1cm} \hat
T_m(x,z) = ((\Lambda^{-1}(z))_m{}^nt_n(x).
\end{eqnarray}
In these expressions $t(x)$ are the corresponding expressions of
$\hat T$ in the $D$-dimensional space-time. Since in the action all
the indices are contracted, the internal dependence vanish.

\subsection{The $D$-dimensional action \label{action}}

Once discussed the general characteristics of the vielbein
parametrization, we apply the $S^3$ group-manifold reduction to the
$(D+3)$-dimensional Hilbert-Einstein action. The important
quantities in the vielbein formalism are the components of the spin
connection ${\hat \omega}_{\hat a \hat b}$. By using the
parametrization (\ref{ansatzmat})-(\ref{nanzatse}), the
$(D+3)$-dimensional spin connection reads
\begin{eqnarray}
 \hat \omega _{ab} & = & \omega_{ab} -2c_1 e^{-c_1\varphi}{\hat
e}_{[a}\partial_{b]}\varphi -\frac{1}{2}e^{(c_2-2c_1)\varphi}
F_{ab}{}^m L_{mn} e^{n}, \nonumber  \\
\hat \omega_{am} & = & (\Lambda^{-1})_m{}^n \left [ e^{c_1\varphi}
e^p \left( c_2
\partial_a \varphi \eta_{pn}+(L^{-1})_{(p}{}^q {\cal D}_a L_{q|n)}
\right) + \frac{1}{2}e^{(c_2-2c_1)\varphi} F_{ab}{}^p L_{pn}
{\hat e}^{b} \right ],  \label{spinconn} \\
\hat \omega_{mn} & = & (\Lambda^{-1})_m{}^p(\Lambda^{-1})_n{}^q
\left[- {\hat e}^a e^{-c_1 \varphi} (L^{-1})_{[p}{}^r{\cal D}_a
L_{r|q]}+e^r e^{-c_2\varphi}\left( {\cal F}_{r[pq]} - \frac{1}{2}
{\cal F}_{pqr}+({\cal R}_r)_{pq}  \right ) \right]. \nonumber
\end{eqnarray}
In these expressions $ F^{m} = 2\partial A^{m} - f_{n
p}{}^{m}A^{n}A^{p} $ is the gauge vector field strength, the scalar
functions  ${\cal F}$ and ${\cal R}$ are defined as
\begin{equation}\label{escfunc}
{\cal F}_{mnp}\equiv (L^{-1})_{m}{}^{q}(L^{-1})_{n}{}^{r}L_{s p}
f_{qr}{}^{s}, \hspace{1cm} ({\cal R}_p)_{mn}\equiv
(L^{-1})_p{}^r(R_r)_{mn},
\end{equation}
whereas the covariant derivative of the internal triad is given by
\begin{equation}
 {\cal D}_\mu L_{m}{}^{n}=\partial_\mu L_{m}{}^{n}-
 A_\mu{}^{p}L_{q}{}^{n}f_{m p}{}^{q} +
 A_\mu{}^{p}L_{m}{}^{q}f_{qp}{}^{n} .
\label{dercovg6}
\end{equation}
Notice that as anticipated, the covariant derivative of the internal
triad reflects its relation with the two Lie algebras under
consideration. The second term corresponds to the standard $SU(2)$
gauging generated by the symmetry of the internal coordinate
transformations whereas the third one is related to the symmetry of
the internal tangent space.

Using the reduced spin connection it turns out that the reduction of
the $(D+3)$-dimensional action is
\begin{equation}
\label{eq:8dtruncaction} S  =  C{\displaystyle\int} d^{D}x
\sqrt{|g|}\,
 \big[ {\cal R}
+{\textstyle\frac{1}{4}}{\rm Tr}\left({\cal D} {\cal M}{\cal
M}^{-1}\right)^{2} +{\textstyle\frac{1}{2}}(\partial \varphi)^2 -
{\textstyle\frac{1}{4}}e^{-\frac{2c_1}{3}(D+1)\varphi}F^{m}{\cal
M}_{m n} F^{n} - {\cal V} \big] \, ,
\end{equation}
where ${\cal V}$ is the scalar potential
\begin{equation}
 {\cal V} ={\textstyle \frac{1}{4}}e^{\frac{2c_1}{3}(D+1)\varphi}\,
             \left[ 2{\cal M}^{m n}
         f_{m p}{}^{q} f_{n q}{}^{p}
+ {\cal M}^{m n}{\cal M}^{p q} {\cal M}_{r s}f_{m p}{}^{r}f_{n
q}{}^{s} \right]\, , \label{gopot}
\end{equation}
and $C$ the group volume defined by $ C(SU(2))= \int d^{\,3}z \,
{\rm det}\, (U_{\alpha}{}^m)=16\pi^2$. In this expression of $C$ we
have used the property that $\det \Lambda=1$. From the covariant
derivative of the internal triad (\ref{dercovg6}), is direct to
compute the covariant derivative of the internal metric ${\cal M}$
\begin{equation}
  {\cal D} {\cal M}_{m n} = \partial {\cal M}_{m n}
 + 2 f_{ q(m}{}^{p}  A^{q} {\cal M}_{n) p} \,,
\label{covderm}
\end{equation}
which reflects its invariant character under transformations in the
internal tangent space.

In conclusion, the two differences produced by apply the new
group-manifold reduction with respect to the standard one are
reflected in the extra term $({\cal R}_p)_{mn}$ of the components
$\hat \omega_{mn}$ of the spin connection (\ref{spinconn}) and in
the extra term in the covariant derivative of the internal triad
(\ref{dercovg6}).

These differences are not manifest in the reduced action and
therefore in the equations of motion either because they are
enterally written in terms of ${\cal M}_{mn}$. The reduced
Lagrangian has the same functional form independently of the
group-manifold reduction applied (either the standard one or the new
one) and therefore the gauge group of the lower-dimensional theory
is either $SO(3)$ or $SU(2)$. The result is expected because the
difference in the parametrization of the vielbein among both
group-manifold reductions is a transformation in the internal
tangent space and the internal metric is invariant under such
transformation. However the new group-manifold reduction has leaved
its imprint in the internal components of the spin connection. In
the next section we shall show a consequence of this result at the
level of the domain wall solutions of the resulting
lower-dimensional theory.

\section{Bianchi type IX domain wall solutions \label{domainw}}

In this section we discuss the domain wall solutions to the
$D$-dimensional action (\ref{eq:8dtruncaction}). The solutions were
originally given for the case $D=1$ and Euclidean signature in
\cite{Gibbons:1979xn}. We shall keep in the following discussion the
generic dimension $D$.

\subsection{The action and the equations of motion \label{secondor}}

After group-manifold reduction the $D$-dimensional field content is
$ \{ e_\mu{}^a, L_{m}{}^n, \varphi, A^{m} \}$. For Bianchi type IX
group manifolds the 5-dimensional scalar coset  $L_{m}{}^n$ contains
two dilatons and three axions. An explicit representation of
$L_m{}^n$ in terms of the five scalars can be found in
\cite{AlonsoAlberca:2003jq,Bergshoeff:2003ri}. In order to simplify
the discussion is convenient to consider the following consistent
truncated parametrization of the scalar coset
\begin{equation}
L_m{}^n(x)=\mbox
{diag}(e^{-\frac{\sigma}{\sqrt{3}}},e^{-\frac{\phi}{2}+\frac{\sigma}{2\sqrt{3}}},
       e^{\frac{\phi}{2}+\frac{\sigma}{2\sqrt{3}}}),
\end{equation}
where we have set the axions to zero. In terms of the dilaton
fields, the action can be rewritten in the following way
\begin{equation}
\label{eq:8daction} S  =  C{\displaystyle\int} d^{D}x \sqrt{|g|}\,
 \big[ {\cal R}
+ {\textstyle\frac{1}{2}}(\partial \phi)^2 +
{\textstyle\frac{1}{2}}(\partial \sigma)^2 +
{\textstyle\frac{1}{2}}(\partial \varphi)^2 -
{\textstyle\frac{1}{4}}e^{-\frac{2c_1}{3}(D+1)\varphi}F^{m}{\cal
M}_{m n} F^{n} - {\cal V} \big] \, ,
\end{equation}
where
\begin{equation}
 {\cal V} = -{\textstyle \frac{1}{4}}e^{\frac{2c_1}{3}(D+1)\varphi}\,
             \left[e^{\frac{2\sigma}{\sqrt{3}}}+ e^{-\phi-\frac{\sigma}{\sqrt{3}}}+e^{\phi-\frac{\sigma}{\sqrt{3}}}
             - e^{-\frac{4\sigma}{\sqrt{3}}}- e^{-2\phi+\frac{2\sigma}{\sqrt{3}}}-e^{2\phi+\frac{2\sigma}{\sqrt{3}}}\right]. \label{8pot}
\end{equation}

We are interested in solutions of cohomogeneity one also known as
domain wall solutions. These are solutions of the theory in the
truncation $A_\mu=0$ that only depend on one spatial coordinate
orthogonal to the compactification manifold, hence  we take the
following ansatz
\begin{equation}
ds^2_D=f^2(y)dx_{(D-1)}^2-g^2(y) dy^2, \label{ansatdw}
\end{equation}
\begin{equation*}
\varphi=\varphi(y), \hspace{0.5cm} L_m{}^{n}=L_m{}^{n}(y).
\end{equation*}
At the beginning due to the ansatz, we have $D+3$ non-trivial second
order equations of motion for the fields, $D$ of them corresponding
to the diagonal components of the metric tensor $g_{\mu \nu}$ and
three corresponding to the scalar fields $\varphi$, $\phi$ and
$\sigma$. However it turns out that only two of the equations of
motion for the metric tensor are independent, the ones for $g_{yy}$
and $g_{00}$ (the other $(D-2)$ for $g_{ii}$ are the same as the
equation of motion for $g_{00}$). It is direct to show that taking
$f(y)=e^{-c_1\varphi}$ the equation of motion for $g_{00}$ becomes
the same as the equation of motion for the scalar field $\varphi$
reducing the system to four independent equations of motion. By the
additional choice $g(y)=e^{(3c_2-c_1)\varphi}$ we can simplify the
equations to the form
\begin{equation}
- e^{2(3c_2-c_1)\varphi} {\cal V}  =   \partial_y^2 \varphi \, ,
\hspace{0.5cm} e^{2(3c_2-c_1)\varphi} \frac{\delta {\cal V}}{\delta
\phi} = \partial_y^2\phi \, , \hspace{0.5cm} e^{2(3c_2-c_1)\varphi}
\frac{\delta {\cal V}}{\delta \sigma} =
\partial_y^2 \sigma \, , \nonumber
\end{equation}
\begin{equation}
- e^{2(3c_2-c_1)\varphi} {\cal V}  =  \frac{10}{6}(\partial_y
\varphi)^2 + \frac{1}{2}(\partial_y \sigma)^2 +
\frac{1}{2}(\partial_y \phi)^2  .
\end{equation}
This system of equations was studied long time ago
\cite{Gibbons:1979xn} and its solutions are well known. In order to
make contact with the original literature we introduce a change of
variables in the following way
\begin{equation}\label{varchange}
\varphi= \ln (abc)^{1/3c_2},\hspace{0.7cm}\sigma= \ln \left(
\frac{bc}{a^2}\right)^{1/\sqrt3},\hspace{0.7cm}\phi=\ln
\left(\frac{c}{b}\right).
\end{equation}
Notice that $a,b$, and $c$ are positive variables. In terms of them
the action reads
\begin{equation}\label{lag4d}
S  \propto {\displaystyle\int} dy\left[ 2\left(\frac{\partial_y
a}{a} \frac{\partial_y b}{b} + \frac{\partial_y b}{b}\frac
{\partial_y c}{c} + \frac{\partial_y c}{c} \frac{\partial_y
a}{a}\right)-\frac{1}{2}(a^4+b^4+c^4-2 a^2 b^2-2 b^2 c^2- 2c^2
a^2)\right ].
\end{equation}
The four equations of motion are
\begin{equation}\label{3emabc}
2\partial_y^2 (\ln a)= a^4-(b^2-c^2)^2 \, ,
\end{equation}
plus the two equations obtained by cyclic permutation of $(a,b,c)$
and
\begin{equation}\label{1emabc}
4\left( \frac{\partial_y a}{a} \frac {\partial_y b}{b} +
\frac{\partial_y b}{b}\frac {\partial_y c}{c} + \frac{\partial_y
c}{c} \frac{\partial_y a}{a} \right ) = 2 a^2 b^2+2 b^2 c^2+ 2c^2
a^2-a^4-b^4-c^4.
\end{equation}
In these variables the $D$-dimensional interval (\ref{ansatdw}) can
be rewritten as
\begin{equation}\label{Dmetric}
ds_D^2=(abc)^{-{2c_1}/3c_2}dx_{(D-1)}^2-(abc)^{(6c_2-2c_1)/3c_2}dy^2,
\end{equation}
and upon uplifting, the $(D+3)$-dimensional interval is of the form
$\mathbb{R}^{D-2,1} \times M_4$, explicitly
\begin{equation}
ds_{D+3}^2=dx_{D-1}^2-((abc)^2dy^2+a^2\sigma_1^2+b^2\sigma_2^2+c^2\sigma_3^2).
\end{equation}
The $D$-dimensional domain wall solutions and the manifolds $M_4$
are entirely given by the three positive functions $a(y)$, $b(y)$
and $c(y)$ satisfying the equations (\ref{3emabc}) and
(\ref{1emabc}). The solutions describe cohomogeneity one self-dual
solutions to the 4-dimensional Euclidean Einstein gravity in empty
space.

We are not going to discuss the whole list of manifold solutions
$M_4$. To our purpose it is enough to mention that some interesting
solutions are the BGPP metrics \cite{Belinskii:1978}, the self-dual
Taub-NUT metrics \cite{Hawking:1977jb,Gibbons:1979xn} and the
Eguchi-Hanson metrics \cite{Eguchi:1978xp,Eguchi:1978gw}.

\subsection{The self-dual spin connection\label{firstor}}

The manifolds $M_4$ are self-dual solutions to the 4-dimensional
Euclidean Einstein gravity. The self-dual character means that for
these manifolds the 4-dimensional curvature is self-dual
($\tilde{R}_{IJ}=R_{IJ}$). It was found that the self-duality
condition of the curvature gives origin to second order differential
equations of motion that accept two different sets of first
integrals \cite{Gibbons:1979xn}. Each set consists of the three
equations
\begin{equation}\label{firstsys}
2\frac{\partial_y a}{a}= -a^2 +  b^2 + c^2-2\lambda bc \, ,
\hspace{2cm} \hbox{and cyclic}.
\end{equation}
If $\lambda=0$ the set of equations is known as the BGPP system
\cite{Belinskii:1978}, whereas if $\lambda=1$, the set of equations
is known as the Atiyah-Hitchin system \cite{Atiyah:1985dv}.

As discussed in \cite{Eguchi:1978gw}, for the 4-dimensional
Euclidean gravity, self-duality in the spin connection is both a
sufficient condition for the self-duality of $R_{IJ}$ and hence for
solving the Einstein equations, and necessary in the sense that if
$R_{IJ}={\tilde R}_{IJ}$ is satisfied, one can always transform
$\omega_{IJ}$ by an $O(4)$ gauge transformation into the form
$\omega_{IJ}= {\tilde \omega}_{IJ}$. The advantage to do this is
that we deal with first order instead of second order differential
equations.

After applying a Bianchi type IX group-manifold reduction to the
$(D+3)$-dimensional spin connection and considering the domain wall
solution (\ref{Dmetric}), we end with six independent non-vanishing
components of the spin connection ($\hat \omega_{\underline ym}$,
$\hat \omega_{mn}$) which correspond to the components of the spin
connection of the manifold $M_4$.

We define the dual of the spin connection as
\begin{equation}
\tilde {\hat \omega}_{IJ}=\frac{1}{2}\varepsilon_{IJ}{}{}^{KL}{\hat
\omega}_{KL},
\end{equation}
where $I,J=\{y,1,2,3\}$ and $\varepsilon_{y123}=1$.

By demanding that the spin connection of the metric $M_4$ in the
basis ($abcdy, a\sigma^1, b\sigma^2,c\sigma^3$) be self-dual we
obtain the BGPP system of first order differential equations
\cite{Belinskii:1978}
\begin{equation}\label{BGPPsys}
2\frac{\partial_y a}{a}= -a^2 +  b^2 + c^2 \, , \hspace{2cm}
\hbox{and cyclic}.
\end{equation}
This parametrization of the basis occurs in the standard
group-manifold reduction. When the three invariant directions are
different, i.e. $a\neq b\neq c$ the equations (\ref{BGPPsys}) admit
the BGPP metrics as solutions \cite{Belinskii:1978} whilst if two of
them are equal i.e. ($a=b\neq c$) admit the Eguchi-Hanson metrics as
solutions \cite{Eguchi:1978xp,Eguchi:1978gw}.

If instead we apply the new $S^3$ group-manifold reduction, by
require a self-dual spin connection we get three independent first
order differential equations
\begin{eqnarray}
c_2\partial_y \varphi-(L^{-1})_{(1}{}^p \partial_y L_{p|1)} &= &
\frac{1}{2}e^{(c_1-c_2)\varphi} (-{\cal F}_{123}+{\cal
F}_{231}-{\cal F}_{312}-2({\cal R}_1)_{23}) , \nonumber \\
c_2\partial_y \varphi-(L^{-1})_{(2}{}^p \partial_y L_{p|2)} &= &
\frac{1}{2}e^{(c_1-c_2)\varphi} (-{\cal F}_{123}-{\cal
F}_{231}+{\cal F}_{312}-2({\cal R}_2)_{31}) , \\
c_2\partial_y \varphi-(L^{-1})_{(3}{}^p \partial_y L_{p|3)} &= &
\frac{1}{2}e^{(c_1-c_2)\varphi} (+{\cal F}_{123}-{\cal
F}_{231}-{\cal F}_{312}-2({\cal R}_3)_{12}) . \nonumber
\end{eqnarray}
Or in terms of the variables $a$, $b$ and $c$ we have, the
Atiyah-Hitchin first order system
\begin{eqnarray}
2\frac{\partial_y a}{a}= -a^2 +  b^2 + c^2 - 2bc \, , \hspace{2cm}
\hbox{and cyclic}. \label{eqdw}
\end{eqnarray}
When two of the tree invariant directions are equal i.e. ($a=b\neq
c$) this system admits the Taub-NUT family of metrics as solutions
\cite{Gibbons:1979xn}. It is important to stress the origin of the
different terms in the r.h.s. of the equation (\ref{eqdw}). The
square terms are associated to the $\cal F$ scalar terms of the
internal spin connection, which come from the structure constants of
the $SU(2)$ Lie algebra (\ref{algk}). The terms like $-2bc$ are
associated to the ${\cal R}$ scalar terms and come from the constant
matrices $R_m$ of the $SO(3)$ Lie algebra (\ref{algr}).

This result should not be surprising, as fact in
\cite{Gibbons:2001ds} was shown that if the 4-dimensional metric is
related to the general class of multi-instantons obtained in
\cite{Gibbons:1978zt}, the self-duality condition in the spin
connection implies that the metric is self-dual Ricci flat.
Depending of the election of a constant parameter, the
multi-instantons become either the multi Taub-Nut metrics or the
multi Eguchi-Hanson metrics. The same result was obtained in the
context of 3-dimensional Toda equations \cite{SfetsosyR}.

Now we have a clear picture of the relation between the two
different Bianchi type IX group-manifold reductions and the domain
wall type solutions of the reduced theory. Because the equations of
motion (second order differential equations) are the same in both
cases, the domain wall solutions coincide as well. However from the
point of view of the first order differential equations, the
solutions are divided into two disjoint sets. One set is given by
the metrics that solve the BGPP system (\ref{BGPPsys}) and the
another one by the metrics that solve the Atiyah-Hitchin system
(\ref{eqdw}). If we reduce pure Einstein gravity applying the
standard group-manifold reduction the domain walls that solve the
BGPP system are self-dual in both the curvature and the spin
connection of $M_4$ whereas the metrics in the another set of
solutions are self-dual only in the curvature. If instead we reduce
applying the ``new" group-manifold reduction the conclusion is the
opposite. The possibility of relate the different first order
systems with the inclusion (or not) of the adjoint matrix $\Lambda$
was already suggested in \cite{Graham:1994qs,Sfetsos:1996pm}.

It is well known that in the case that $a,b$ and $c$ are positive
variables, one of the Eguchi-Hanson metrics and one of the Taub-NUT
metrics are the only complete non-singular $SO(3)$ hyper-K\"ahler
metrics in four dimensions \cite{Gibbons:1979xn,Bakas:1996gf}, both
of them are obtained in the case in which two of the invariant
directions are equal. From the $(D+3)$-dimensional point of view
these two solutions correspond to $\mathbb{R}^{D-2,1} \times M_4$
with either $M_4$ the Eguchi-Hanson metric \cite{Eguchi:1978xp}
whose generic orbits are $\mathbb{R}P^3$ \cite{Belinskii:1978} or
the self-dual Taub-NUT solution whose generic orbits are $S^3$
\cite{Hawking:1977jb}. In the latter case, the complete
$(D+3)$-metric is known as the Kaluza-Klein monopole
\cite{Sorkin:1983ns,Gross:1983hb}.

\subsection{First order equations and the
superpotential\label{supot}}

As established in \cite{Gibbons:2001ds}, the Lagrangian of the
action (\ref{lag4d}) can be written as
\begin{equation}
L=T-V=\frac{1}{2}g_{mn}\left ( \frac{\partial \alpha_m}{\partial
y}\right )\left( \frac{\partial \alpha_n}{\partial y} \right) +
\frac{1}{2}g^{mn}\left( \frac{\partial W}{\partial \alpha_m}\right)
\left( \frac{\partial W}{\partial \alpha_n} \right),
\end{equation}
where $\alpha_m \equiv (\ln a, \ln b, \ln c)$ and $W$ is a
superpotential given by
\begin{equation}
W=a^2+b^2+c^2-2\lambda bc -2 \lambda ca -2\lambda ab.
\end{equation}
We conclude that the case $\lambda=0$ is related to the standard
group-manifold reduction whereas the case $\lambda=1$ is related to
the new group-manifold reduction. In the literature concerning
domain wall solutions is usual to write down the superpotential in
terms of the original variables, i.e. in terms of the dilatons. The
inverse variable transformations of (\ref{varchange}) are
\begin{equation}
a(y) \equiv e^{c_2\varphi - \frac{\sigma}{\sqrt{3}}}, \hspace{0.5cm}
b (y) \equiv e^{c_2\varphi +
\frac{\sigma}{2\sqrt{3}}-\frac{\phi}{2}}, \hspace{0.5cm} c (y)\equiv
e^{c_2\varphi + \frac{\sigma}{2\sqrt{3}}+\frac{\phi}{2}}.
\end{equation}
It is straightforward to show that in terms of the superpotential
and the dilatons, the potential satisfies the property
\begin{equation}
V=\frac{1}{2}\left( \left(\frac {\partial W}{\partial
\varphi}\right)^2+\left(\frac {\partial W}{\partial
\phi}\right)^2+\left(\frac {\partial W}{\partial \sigma}\right)^2
-\left(\frac{D-1}{D+1}\right)W^2 \right).
\end{equation}
It is also possible to write down the BGPP first order system
(\ref{BGPPsys}) and the Atiyah-Hitchin first order system
(\ref{eqdw}) in terms of the dilatons and the superpotential. The
equations in this case become
\begin{equation}
\frac{\partial \varphi}{\partial y}=\frac{1}{6c_2} W , \hspace{1cm}
\frac{\partial \phi}{\partial y}=-\frac{\partial W}{\partial \phi},
\hspace{1cm} \frac{\partial \sigma}{\partial y}=-\frac{\partial
W}{\partial \sigma},
\end{equation}
which are related with first-order Bogomol'nyi equations (see for
example
\cite{Cvetic:1991vp,Skenderis:1999mm,Bakas:1999fa,Bergshoeff:2004nq}
and references therein).

\section{Conclusions\label{conclusions}}

In this paper we have introduced a new consistent $S^3$
group-manifold reduction of Einstein pure gravity in the vielbein
formulation by considering the two 3-dimensional Lie algebras that
$S^3$ admits. We have showed that although the lower-dimensional
theory has the same Lagrangian independently of the Bianchi type IX
group-manifold reduction used (either the standard one or the new
one), there exist two differences produced by apply the new
group-manifold reduction with respect to the standard one. These
differences are a) A new term in the components of the spin
connection with two internal indices and b) an additional term in
the covariant derivative of the internal triad.

As an application of these results we studied the domain wall
solutions of the lower-dimensional theory and we concluded that
because the equations of motion coincide for both Bianchi type IX
group-manifold reductions, the domain wall solutions coincide as
well. Upon uplifting to the higher-dimension these are purely
gravitational solutions of the form $\mathbb{R}^{D-2,1} \times M_4$.
However from the point of view of first order differential
equations, the solutions are divided into two disjoints sets, one
set is given by the metrics that solve the BGPP system and the
another set by the metrics that solve the Atiyah-Hitchin system. If
we reduce pure Einstein gravity applying the standard group-manifold
reduction the domain walls that solve the BGPP system are self-dual
in both the curvature and the spin connection of $M_4$ whereas the
metrics in the another set of solutions are self-dual only in the
curvature. If instead we reduce applying the new group-manifold
reduction the conclusion is the opposite, i.e. now the domain walls
that solve the Atiyah-Hitchin system are self-dual in both the
curvature and the spin connection of $M_4$  whilst metrics in the
another set of solutions are self-dual only in the curvature. This
result can be relevant if we are working with a formulation of
gravity in which the fundamental field is the vielbein instead of
the metric, for instance, in supergravity.

We believe the results of this paper open the possibility to
construct an 8-dimensional gauged supergravity by apply the new
$S^3$ group-manifold reduction to the 11-dimensional supergravity.
The hope is that this supergravity could admit a 1/2 BPS domain wall
solution which upon uplifting to eleven dimensions led to the
Kaluza-Klein monopole. This problem should be analogous to the
results obtained in the context of the 4-dimensional $N=1$
supergravity \cite{D'Eath:1993up,Graham:1994qs}. In this case the
non-singular solution $\mathbb{R}^{6,1} \times M_4$ of the
11-dimensional supergravity should be the analogous of the wormhole
state if $M_4$ is the Eguchi-Hanson metric \cite{Eguchi:1978xp} and
of the Hartle-Hawking state if $M_4$ is the self-dual Taub-NUT
solution \cite{Hawking:1977jb}. It would be interesting to establish
this analogy explicitly.

Usually when the internal orthonormal frame specified by $e^m$ is
rotated by a $z$-dependent orthogonal transformation
${e}^m\rightarrow e^n \lambda_n{}^m(z)$, the change in the
corresponding spin connection is interpreted as a gauge
transformation because the spin connection transforms exactly like a
Yang-Mills potential \cite{Eguchi:1978gw}. It would be interesting
to see whether the local adjoint matrix $\Lambda(z)$ can be
interpreted as a kind of large gauge transformation, at least for
some special type of solutions. This interpretation is suggested
because the adjoint matrix changes the expression of the spin
connection in such a way that when we impose the condition of
self-duality, we obtain two different systems of first order
differential equations that accept solutions with isometry groups of
different topology.

The tools used in this paper are not exclusive to 3-dimensional
group-manifolds and it would be interesting to see whether the
generalization to other dimensions is possible \cite{Jantzen}.

\acknowledgments

\bigskip

We are grateful to Eric Bergshoeff, Ulf Gran, Hugo Morales, Mikkel
Nielsen, Tom\'as Ort\'{\i}n and Diederik Roest for interesting and
useful discussions at different stages of this work. In particular
we should like to thank Mikkel Nielsen, Tom\'as Ort\'{\i}n and
Diederik Roest for a critic reading to an early version of the
draft. Special thanks to Emily and Pepe for encourage me to work in
this project each time it had gone. We also would like to thank the
Centre for Theoretical Physics  at University of Groningen and the
Instituto de Ciencias Nucleares at Universidad Nacional Aut\'onoma
de M\'exico for hospitality and financial support at different
stages of this work. This research was supported in part by the
project CONACyT-40745-F.

\appendix

\section{Bianchi type IX Lie groups}\label{Lg3d}

In the next discussion we give explicit expressions for the relevant
quantities used in the $S^3$ group-manifold reduction, we follow the
conventions of \cite{Jantzen}. The starting point is to assume a
3-dimensional vector fields basis ${\bf K}_m$ that satisfies the Lie
algebra $\mathfrak{g}_3$

\begin{equation}
[{\bf K}_m,{\bf K}_n]=f_{mn}{}^{p}{\bf K}_p .
\end{equation}

In the Bianchi type IX case the expression for the structure
constants can be diagonalized and taken as

\begin{equation}
 f_{m n}{}^{p}= \epsilon_{m n q}
 \delta^{q p}, \hspace{1cm}
 \delta^{m n} = {\rm diag}(1,1,1) \ .
\end{equation}

Choosing the matrices $\{ {\bf e}_m{}^n \}$ as the basis of
$\mathfrak{sl}(3, \mathbb{R})$ where ${\bf e}_m{}^n$ is the matrix
whose only non-vanishing component is a one in the $m^{\mbox{th}}$
row and $n^{\mbox{th}}$ column, the canonical basis $\{R_m \}$ of
the canonical adjoint group $Ad_{\bf K}(G)$ is defined as the
adjoint representation of the generators ${\bf K}$ in this basis,
$(R_m)=f_{mn}{}^p {\bf e}_p{}^n$, explicitly

\begin{equation}
(R_1)_m{}^n  = \left(
\begin{array}{ccc}
0 &   0   & \, \, \, 0    \\
0 &   0   & -1  \\
0 &   1   & \, \, \, 0
\end{array}
\right) \, , \hspace{0.5cm} (R_2)_m{}^n   = \left(
\begin{array}{ccc}
\, \, \, 0  &  0   &  1 \\
\, \, \, 0  &  0   &  0 \\
        -1  &  0   &  0
\end{array}
\right) \, , \hspace{0.5cm} (R_3)_m{}^n  = \left(
\begin{array}{ccc}
 0  &          -1  & 0  \\
 1  &  \, \, \, 0  & 0  \\
 0  &  \, \, \, 0  & 0
\end{array}
\right) \, ,
\end{equation}
and satisfy the algebra
\begin{equation}
[R_m,R_n]= f_{mn}{}^p R_p.
\end{equation}
 Exponentiating the generators of the Lie algebra
$\mathfrak{g}_3$ in the adjoint representation, we get the adjoint
representation $\Lambda(z)$ of the group $G_3$

\begin{eqnarray}
\Lambda_m{}^n(z) & = &  e^{z^1 R_1} e^{z^2 R_2} e^{z^3 R_3} = \left(
\begin{array}{ccc}
 c_{2} c_{3}   &  -c_{2}s_{3}    &  s_{2}           \\
 c_{1}s_{3}+c_{3}s_{1}s_{2}  &
 c_{3}c_{1}-s_{1}s_{2}s_{3}  & -s_{1}c_{2}   \\
 s_{1}s_{3}-c_{3}c_{1}s_{2}  &
 c_{3}s_{1}+c_{1}s_{2}s_{3}  & c_{1}c_{2}
\end{array}
\right ) ,
\end{eqnarray}

\noindent where we have used the following abbreviations $(a=1,2,3)$

\begin{equation}
c_{a} \equiv \cos z^a, \hspace{1cm} s_{a} \equiv \sin z^a.
\end{equation}

It can be directly checked that $\det \Lambda = 1$ and also that the
matrix is {\it orthogonal} $\Lambda_{m}{}^p(z) \Lambda_n{}^q(z) \
\eta_{pq}=\eta_{mn}$. The next step is to compute the left invariant
1-forms using the equation ${\bf \Lambda}^{-1}d{\bf \Lambda} =
\sigma^m R_m$. Its dual base $\{ {\bf K}_m \}$ can also be obtained
by require $\sigma^m {\bf K}_n=\delta_m{}^n$.

\begin{eqnarray}
\sigma^1 & = & \cos z^2 \cos z^3 dz^1 + \sin z^3 dz^2,
\hspace{0.8cm}
{\bf K}_1=\frac{\cos z^3}{\cos z^2} \partial_1 + \sin z^3 \partial_2 - \frac{\cos z^3 \sin z^2}{\cos z^2} \partial_3 \, ,\nonumber \\
\sigma^2 & = & -\cos z^2 \sin z^3 dz^1 + \cos z^3 dz^2,
\hspace{0.4cm}
{\bf K}_2=-\frac{\sin z^3}{\cos z^2} \partial_1 + \cos z^3 \partial_2 + \frac{\sin z^3 \sin z^2}{\cos z^2} \partial_3 \, ,\nonumber \\
\sigma^3 & = & \sin z^2 dz^1 + dz^3,\hspace{2.9cm} {\bf
K}_3=\partial_3 \, .
\end{eqnarray}

From these expressions we have that the matrix $U_\alpha{}^m(z)$ is
given by

\begin{equation}
U_\alpha{}^m (z) = \left(
\begin{array}{ccc}
 \cos z^2 \cos z^3 &  -\cos z^2 \sin z^3 &  \sin z^2 \\
      \sin z^3     &       \cos z^3       &     0     \\
         0         &          0          &     1
\end{array}
\right ) .
\end{equation}

The relation between the left and right invariant Lie algebras, and
the relation between the left and right invariant 1-forms is

\begin{equation}
{\bf \tilde K}_m = \Lambda_m{}^n {\bf K}_n , \hspace{2cm} {\tilde
\sigma}^m = \sigma^n (\Lambda^{-1})_n{}^m.
\end{equation}

Using them  we get

\begin{eqnarray}
\tilde \sigma^1 & = &  \sin z^2 dz^3 + dz^1,
\hspace{2.9cm} {\bf \tilde K}_1= \partial_1 \, ,\\
\tilde \sigma^2 & = & - \cos z^2 \sin z^1 dz^3 + \cos z^1 dz^2 ,
\hspace{0.4cm} {\bf \tilde K}_2= - \frac{\sin z^1}{\cos z^2}
\partial_3 + \cos z^1 \partial_2 + \frac{\sin z^1 \sin z^2}{\cos z^2} \partial_1 \, ,\nonumber \\
\tilde \sigma^3 & = & \cos z^2 \cos z^1 dz^3 + \sin z^1 dz^2 ,
\hspace{0.8cm} {\bf \tilde K}_3= \frac{\cos z^1}{\cos z^2}
\partial_3 + \sin z^1 \partial_2 -\frac{\cos z^1 \sin z^2}{\cos z^2}
\partial_1 \, . \nonumber
\end{eqnarray}

All these quantities satisfy the Lie algebra $\mathfrak{g}$ ($\tilde
{\mathfrak{g}}$)
\begin{equation}
[{\bf K}_m,{\bf K}_n]=f_{mn}{}^{p}{\bf K}_p, \hspace{1cm} [\tilde
{\bf K}_m,\tilde {\bf K}_n]=-f_{mn}{}^{p}\tilde {\bf K}_p,
\hspace{1cm} [{\bf K}_m,\tilde {\bf K}_n]=0  .
\end{equation}
and the Maurer-Cartan equations
\begin{equation}
d{\bf \sigma}^m=-\frac{1}{2}f_{np}{}^{m}{\bf \sigma}^n \wedge {\bf
\sigma}^p , \hspace{1cm} d \tilde {\bf
\sigma}^m=\frac{1}{2}f_{np}{}^{m}\tilde {\bf \sigma}^n \wedge \tilde
{\bf \sigma}^p.
\end{equation}

\end{document}